\documentclass[conference]{IEEEtran}
\IEEEoverridecommandlockouts
\usepackage{cite}
\usepackage{amsmath,amssymb,amsfonts}
\usepackage{algorithm}
\usepackage{algorithmic}
\usepackage{graphicx}
\usepackage{textcomp}
\usepackage{xcolor}
\usepackage{xspace}
\usepackage{soul}
\usepackage{kotex} 

\def\BibTeX{{\rm B\kern-.05em{\sc i\kern-.025em b}\kern-.08em
    T\kern-.1667em\lower.7ex\hbox{E}\kern-.125emX}}
\begin{document}

\title{Trends in LEO Satellite Handover Algorithms}

\author{\IEEEauthorblockN{Soohyun Park and Joongheon Kim}
\IEEEauthorblockA{Department of Electrical and Computer Engineering, Korea University, Seoul, Republic of Korea 
\\
E-mails: \texttt{soohyun828@korea.ac.kr}, 
\texttt{joongheon@korea.ac.kr}
}
}
\maketitle

\begin{abstract}
In this paper, we review well-known handovers algorithms in satellite environment. The modern research trends and contributions are proposed and summarized in order to overcome their considering problems in satellite-air-ground integrated network environment caused by the fast movement of Low Earth Orbit (LEO) satellite and related frequent handover occurrences. 
\end{abstract}

\section{Introduction}
\begin{figure}[t]
    \begin{center}
        \includegraphics[width=0.95\linewidth]{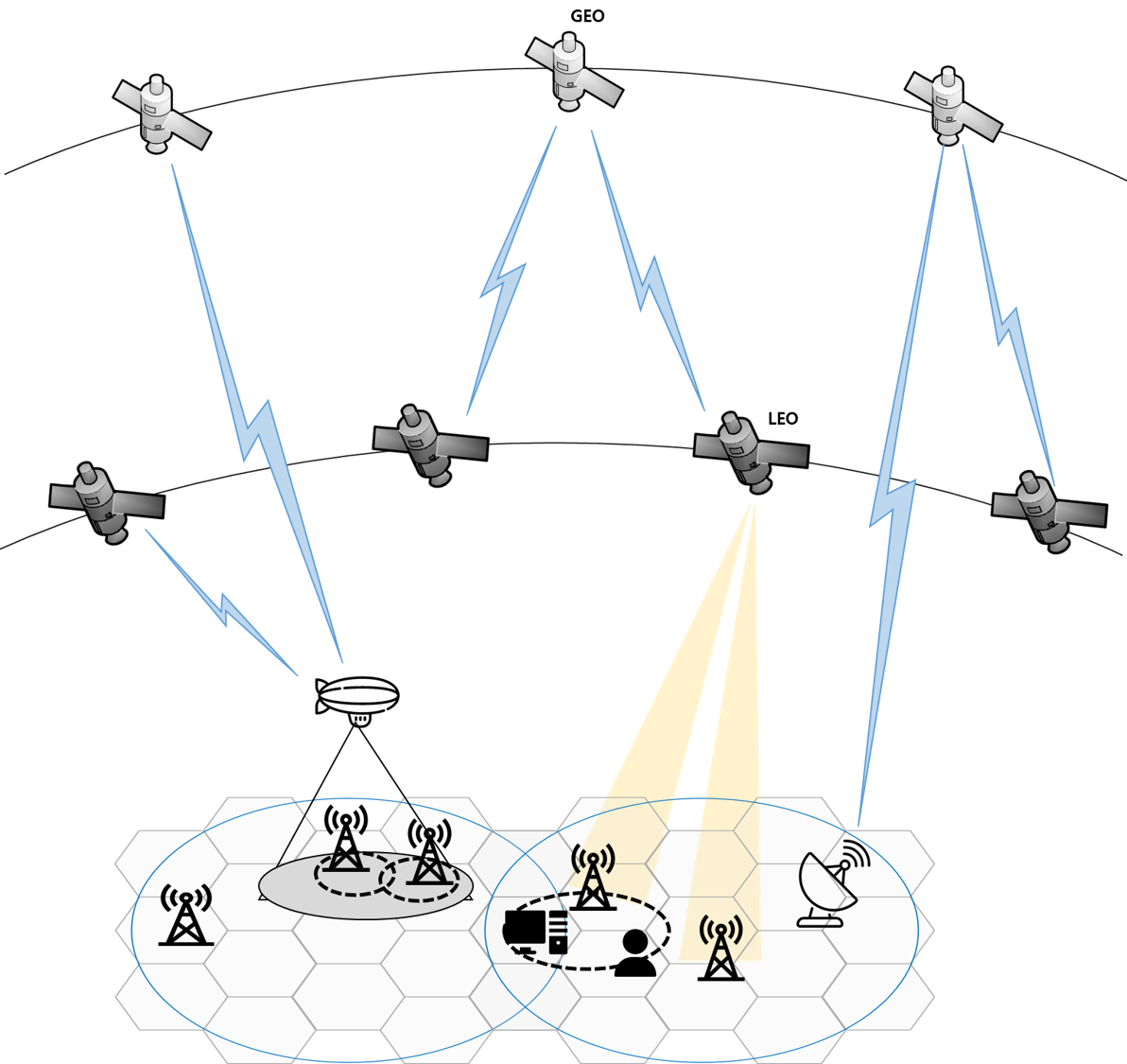}
    \end{center}
    \caption{A satellite integrated network architecture.}
    \label{fig:network}
\end{figure}

With the development of terrestrial networks based on 5G communication networks and the advent of next generation 6G communication networks, the mobile satellite system where various types of satellites exist all together as shown in Fig.~\ref{fig:network}, has become one of major research topics, recently.
The satellite, which can be considered as one of major components in newly discussed integrated networks in 6G research, can be classified as geostationary earth orbit (GEO), medium earth orbit (MEO), and low earth orbit (LEO) satellites, depending on the heights of the existing elevations and the forms of orbits~\cite{Comm_Survey2021}.
Among the various types of satellites, LEO satellites that are deployed at altitudes between $500$ and $2,000$\,km are used in various communication and network applications in recent research related to global world-wide communications.
Along with the research trends using LEO, there are several global commercial deployments such as SpaceX, Kepler, Telesat, and Starlink constellations that consist of the lowest $140$ satellites or up to $42,000$ LEO satellites {\cite{LEO3,Comm_Survey2021,apwcs21kim}}.

LEO satellites that have the closest physical distance from the ground provide low delays, low energy consumption and efficient frequency spectrum utilization compare to other types of satellites, e.g., GEO and MEO~\cite{LEO1}.
However, the high-speed movements of LEO satellites make handover decisions. In addition, the small ground coverage of LEO satellites is also a reason for frequent handover occurrences~\cite{LEO2}. The frequent handovers definitely make impacts on the quality of service (QoS) in wireless communications and networks. 
For these reasons, a new efficient management system is required to maximize user QoS and to manage LEO satellite system communication resources efficiently. 
Some research results are already processed in various manners to overcome the disadvantage of frequent handover occurrences~\cite{Comm_Survey2021, GLOBECOM2019, ICCC2020}.

The main objective of this paper is about to provide and summarize the trends on several significant research results for handover decision control in satellite integrated networks.

The rest of this paper is organized as follows. 
Sec.~\ref{sec:sec2} overviews and classifies handover research trends in satellite networks. 
Sec.~\ref{sec:sec3} presents various research results for handover decision algorithms. 
In addition, Sec.~\ref{sec:sec4} presents future research directions. 
Lastly, Sec.~\ref{sec:sec5} concludes this paper and then presents future work directions.

\section{Handover Mechanisms in Mobile Satellite Networks}\label{sec:sec2}
\subsection{Link-Layer Handover}
Link-layer handovers occur when one or more links change between the communication endpoints, such as satellites or users. 
It is influenced by dynamic connectivity patterns of LEO satellites and is classified as follows~\cite{Comm_Survey2006}.

\begin{itemize}
    \item \textit{Spot-beam handover} is also known as intra-satellite handover. When the end-user leaves the allocated spot-beam coverage, the satellite associated with the end-user does not change and the end-user is reallocated to another spot-beam. It happens frequently in 1-2min intervals due to the small sizes of spot-beam coverage~\cite{Comm_Survey2006}.
    
    \item \textit{Satellite handover} is also known as inter-satellite handover. It occurs when the end-user is transferred from the previously connected satellite to another satellite. 
    
    \item \textit{Inter-satellite link (ISL) handover} happens when interplane inter-satellite links (ISLs) is temporarily switched off due to the change in terms of distances and viewing angles between satellites in neighbor orbits. Then the ongoing connections using these ISL links have to be rerouted, causing ISL handovers.
\end{itemize}

There are two indicators in trade-off relationships to evaluate the performance of link-layer handovers. One is call blocking probability $P_b$ which represents the probability of a new call being blocked during handover. The other one is forced termination probability $P_f$ where the meaning is the probability of a handover call being blocked during handover~\cite{Comm_Survey2006, Comm_Survey2021}. 

\subsection{Network-Layer Handover}
When communication endpoints change Internet protocol (IP) address, the change of coverage areas of satellite or the mobility of the user terminal let Network-layer handover handling occurs~\cite{Comm_Survey2006}. 

\begin{itemize}
    \item \textit{Hard-handover schemes} occur when the current link is released before the next link is established.
    
    \item \textit{Soft-handover Schemes} occur when the current link will not be released until the next connection is established.
    
    \item \textit{Signaling-diversity Schemes} are similar to soft handover, signaling flows through both old and new links and the user data go through the old link during handover.
\end{itemize}

\section{Handover Decision Algorithms using Various Decision Criteria}\label{sec:sec3}
In this section, we overview prior result results in terms of satellite handover control.
Each research result is described from the points of how handover problems are formalized and which solutions are proposed to achieve what objectives in the satellite network environment.

In following subsections, various criteria are presented in terms of reinforcement learning (refer to Sec.~\ref{sec:sec3-1}), game theory (refer to Sec.~\ref{sec:sec3-2}), and optimization criteria (refer to Sec.~\ref{sec:sec3-3}).

\subsection{Reinforcement Learning}\label{sec:sec3-1}
The literature assumes that terminal users can obtain only partial information from satellites without a central controller, and the satellites have limited channel budgets~\cite{GLOBECOM2020}. As a result, competition among satellites occurs.
In this situation, the proposed algorithm in~\cite{GLOBECOM2020} aims to minimize average satellite handovers while satisfying the load constraints of each satellite.

In~\cite{GLOBECOM2020}, the relationship between user and satellite is represented as a graph, and the graph updates when the satellite moves out of or enters the user's sight. The proposed algorithm in~\cite{GLOBECOM2020} set variables $c$ and $x$ where each one means whether the user exists in the coverage of the satellite and whether the user is served bu the satellite. An optimization formulation for service association indicators is formulated and the problem is also for improving channel utilization efficiency in the LEO satellite network simultaneously~\cite{GLOBECOM2020}. 
However, the optimization formula which is a combinatorial integer optimization problem corresponds to NP-hard. To solve the NP-hard problem, the proposed algorithm in~\cite{GLOBECOM2020} transforms the problem into an Multi-Agent Reinforcement Learning (MARL)-based optimization framework. 

For the proposed MARL-based approach, the users in the satellite network system become agents, the settings of state, action, reward are as follows.
\begin{itemize}
    \item \textit{State:} Each agent's status information which includes (i) whether the agent exists in the coverage of the satellite, (ii) the available channels of the satellites, and (iii) the remaining visible time of the satellite.
    \item \textit{Action:} Whether the user is served by satellite at $t$. Note that the value of the service association indicator $x$.
    \item \textit{Reward:} There are three types of reward values depending on the state condition.
\end{itemize}

The simulation-based performance evaluation results show that the proposed MARL-based handover control algorithm is able to reduce not only the average numbers of handover occurrences but also the user blocking rate $P_b$. As a result, it improves the satellite channel utilization of the entire system~\cite{GLOBECOM2020}.

\subsection{Game Theory}\label{sec:sec3-2}
The goal of this research is to decrease handover time/latency while using the network resource efficiently~\cite{Access_HO_game}. The the proposed algorithm in~\cite{Access_HO_game} assumes that the satellite-based network has a software-defined satellite network (SDSN) architecture and models the handover of the network as a bipartite graph. 
In the network system, there are many terminal devices, LEO satellites, and the multiple mobile terminals. Then, they compete for satellite resources and available channels.

To achieve the objective, two sub-algorithms are proposed. One is a handover algorithm based on potential games that maximizes the benefits of mobile devices. The other is a terminal random-access algorithm that makes the satellite network workload be balanced ~\cite{Access_HO_game}. 
Especially, in the handover decision algorithm, the system reaches Nash equilibrium. 
The proposed algorithm in~\cite{Access_HO_game} set an utility function which contains two kinds of functions, i.e., (i) gain function and (ii) loss function, in the course of the application of the potential game theory. 
Each of the functions generates utility values based on (i) visible time and the satellite elevation angle and (ii) handover request time and response time. 
Through the game theory based approach, the average number of handovers in the network is significantly reduced and the call quality of users is also improved. 

\subsection{Optimization Criteria}\label{sec:sec3-3}
As mentioned above, the fast movement of LEO satellites causes frequent handovers to users and the frequent handover occurrences become difficult to guarantee QoS requirements.
In~\cite{Access_HO_forecast}, the network environment assumes that several mobile terminals (MTs), high-altitude platforms (HAPs), LEO, and GEO satellites exist as a multi-layer structure and cross-layer handovers are able to arise among the different levels of layers~\cite{Access_HO_forecast}. The proposed algorithm in~\cite{Access_HO_forecast} researches how to reduce the dropping probability and also to increase the throughput in the network environment.

In~\cite{Access_HO_forecast}, the problem solution is approached in the direction of considering user priority, minimum requirement, delay requirement, channel gain, and the data traffic of beams. An utility function which contains the above items is designed and the handover problem is divided into three parts such as (i) user association, (ii) time slot, and (iii) frequency resource and power allocation~\cite{Access_HO_forecast}. Each part of the handover problems converts to individual optimization problems. The optimal solution of the handover problem is derived by applying Lagrange dual methods and Karush-Kuhn-Tucker (KKT) conditions. 
As a result, the advantage of the proposed method is provided by showing better performance not only on delay and signalling cost compared with traditional handover protocols, but also on dropping probability and throughput compared with other methods. 

\section{Potential Future Trends}\label{sec:sec4}
For our future research directions, we can consider following aspects.
\begin{itemize}
    \item \textit{Video/contents-specific characteristics:} For designing handover decision algorithms, contents-aware decision functions are desired in order to increases user QoS. Thus, video/contents-related characteristics should be essentially considered~\cite{ton201608kim,jsac201806choi,tmc201907koo,twc201912choi,mm2017koo,twc202012choi,twc202104choi}. 
    \item \textit{Unmanned aerial vehicle (UAV)-codesign related characteristics:} Nowadays, there are many research results in satellite-air-ground integrated networks (SAGIN). Thus, the handover decision algorithms should consider UAV-related characteristics~\cite{tvt202106jung,tvt201905shin}.
    \item \textit{Millimeter-wave channel characteristics:} For establishing fast communications within SAGIN networks, high-speed networking should be realized by utilizing millimeter-wave links~\cite{tvt2021jung,jcn201310kim}. Thus, the millimeter-wave wireless technology related characteristics should be considered. 
    \item \textit{Caching characteristics:} The caching functionalities can be implemented in next-generation satellite systems~\cite{tmc202106malik}. The corresponding characteristics can be useful for handover decisions. Thus, it can be considered as well. 
    \item \textit{Mobile device characteristics:} In modern mobile platforms, the computing capabilities are increased a lot. Thus, the use of various high-performance computing in mobile smartphones are available. Therefore, handover decision algorithm design and implementation based on the characteristics can be helpful for better performance and better user QoS~\cite{tmc2021yi,mobisys2010paek,pieee202105park,icdcs2020meteriz}.
\end{itemize}

\section{Conclusions and Future Work}\label{sec:sec5}
In the integrated networks that combine satellite, especially LEO, the design and implementation of handover decision algorithm is definitely essential even in the environment where user mobility is ignored by LEO's characteristics, such as the fast movement rate of LEO and the size of beam coverage.
The handover decision algorithms allow the connection for new satellites to provide continuous services to users even if the connection between the existing satellite and the user is expired.
However, according to fact that the frequent handover reduces user QoS and creates additional drawbacks, proper handover decision algorithms are desired.
Therefore, in this paper, we explore possible types of handovers in satellite environment, and handover control in various network scenarios. Moreover, we confirm that each object is achieved through different approaches.
Based on prior research examples that address the handover problems that need to be addressed to increase communication efficiency and user QoS, new approaches can proposed.

\section*{Acknowledgment}
This research is supported by the National Research Foundation of Korea (NRF-Korea, 2021R1A4A1030775). J. Kim is a corresponding author of this paper.

\bibliographystyle{IEEEtran}
\bibliography{ref_aimlab,ref_handover}

\end{document}